\documentclass[11pt,twoside]{article}
\usepackage{asp2010}

\resetcounters

\begin{document}

\title{The mass-loss rates and molecular abundances of S-type AGB stars}
\author{Sofia~Ramstedt$^{1,2}$, Fredrik~L.~Sch\"oier$^3$, and Hans~Olofsson$^{1,3}$
\affil{$^1$Department of Astronomy, Stockholm University, SE-10691 Stockholm, Sweden}
\affil{$^2$Argelander Institut f\"ur Astronomie, Auf dem H\"ugel 71, DE-53121 Bonn, Germany}
\affil{$^3$Onsala Space Observatory, SE-43992 Onsala, Sweden}}

\begin{abstract}
The S-type stars are believed to have a C/O-ratio close to unity (within a few percent). They are considered to represent an intermediate evolutionary stage as AGB stars evolve from oxygen-rich M-type stars into carbon stars. As possible transition objects the S-type stars could give important clues to the mass-loss mechanism(s) and to the chemical evolution along the AGB. Using observations of circumstellar radio line emission in combination with a detailed radiative transfer analysis, we have estimated mass-loss rates and abundances of chemically important molecules (SiO, HCN) for a sample of 40 S-type AGB stars. The results will be compared to previous results for M-type and carbon stars.
\end{abstract}

\section{Why Care About the S-type AGB stars?}
Earlier theoretical results suggest that the S-type AGB stars have a C/O-ratio$\approx$1 \citep{scalross76}. This has been interpreted as the S-stars being representative of a brief transitional phase as M-type stars, through the dredge-up of internally synthesized carbon, evolves into carbon stars. As such, their study could potentially give important clues to a number unsolved issues regarding the mass-loss mechanism(s) and the chemical evolution as stars evolve along the AGB. In an atmosphere where the chemistry is in equilibrium and the amount of carbon is approximately equal to the amount of oxygen, almost all of the carbon and oxygen will be bound in CO molecules and little will be left to form other carbon- or oxygen-bearing compounds. This complicates the formation of dust in these stars as either free carbon or oxygen in the form of silicon oxide is normally needed to form dust. This does not necessarily imply that dust cannot be formed around these stars, merely that it is more complicated than previously assumed.

The S-type stars are classified by the presence and dominance of ZrO-bands in their optical spectra. Intrinsic S-type stars (as opposed to extrinsic S-type stars which owe their chemical peculiarities to mass-transfer across a binary system) also show Tc in their spectra. Recent models have shown that this classification might represent a larger range in C/O-ratio, possibly going as low as 0.5 (see Van Eck et al., this volume). The stars in our sample are stars found in the {\em General Catalogue of Galactic S stars}, the IRAS {\em Point Source Catalogue}, and the {\em Guide Star Catalogue}. They are all intrinsic and previously detected in CO. This might introduce a bias toward higher mass-loss-rate stars, and we will miss any very low mass-loss-rate stars, however this selection was made with the aim of also detecting line emission from other molecules besides CO. The sample is most likely representative of mass-losing S-type stars and complete (or close to complete) out to 600\,pc \citep{ramsetal09}.

The goals of this investigation is to determine reliable mass-loss rates and circumstellar molecular abundances for chemically important molecules in a consistent way so that the results can be compared to previous results on M-type and carbon stars, hoping that this will lead to a better understanding of the evolutionary status of the S-type stars and the chemical evolution along the AGB.

\section{Observations and modeling}
The 40 sample stars are observed and detected in several lower-frequency rotational transitions of CO ($J=1\rightarrow0$ to $J=3\rightarrow2$; 40 detected sources), SiO ($J=2\rightarrow1$ to $J=8\rightarrow7$; 26 detected sources), and HCN ($J=1\rightarrow0$ to $J=4\rightarrow3$; 18 detected sources) using the Onsala, IRAM, APEX and JCMT telescopes \citep[see][for details on the observations]{ramsetal09,schoetal10}. We have also searched for line emission from SiS, CS, and H$_{2}$CO, but with no or little success.

The data is then modeled using a detailed, non-local, non-LTE, Monte-Carlo radiative transfer code described in e.g. \citet{schoolof01}. The circumstellar envelope is assumed to be spherically symmetric and formed by a constant mass-loss rate. It is also assumed to have a constant expansion velocity. The thermal balance is solved self-consistently. The mass-loss rates ($\dot{M}$) and the kinetic gas temperature distributions ($T(r)$) are determined by reproducing the CO line emission. These parameters are then used as input together with the fractional abundance (relative to H$_{2}$) of the respective molecules in order to reproduce the SiO and HCN line emission (see Figs~1 and 2). The abundance distribution is described by a Gaussian function, and in the modeling both the molecular abundance at the inner boundary ($f_{0}$) and the extent of the emitting region (r$_{\rm{e}}$) is constrained. For all models $\chi^2$-minimization is used to find the best fit.

\begin{figure}[hb]
\begin{flushleft}
\includegraphics[width=3.2cm]{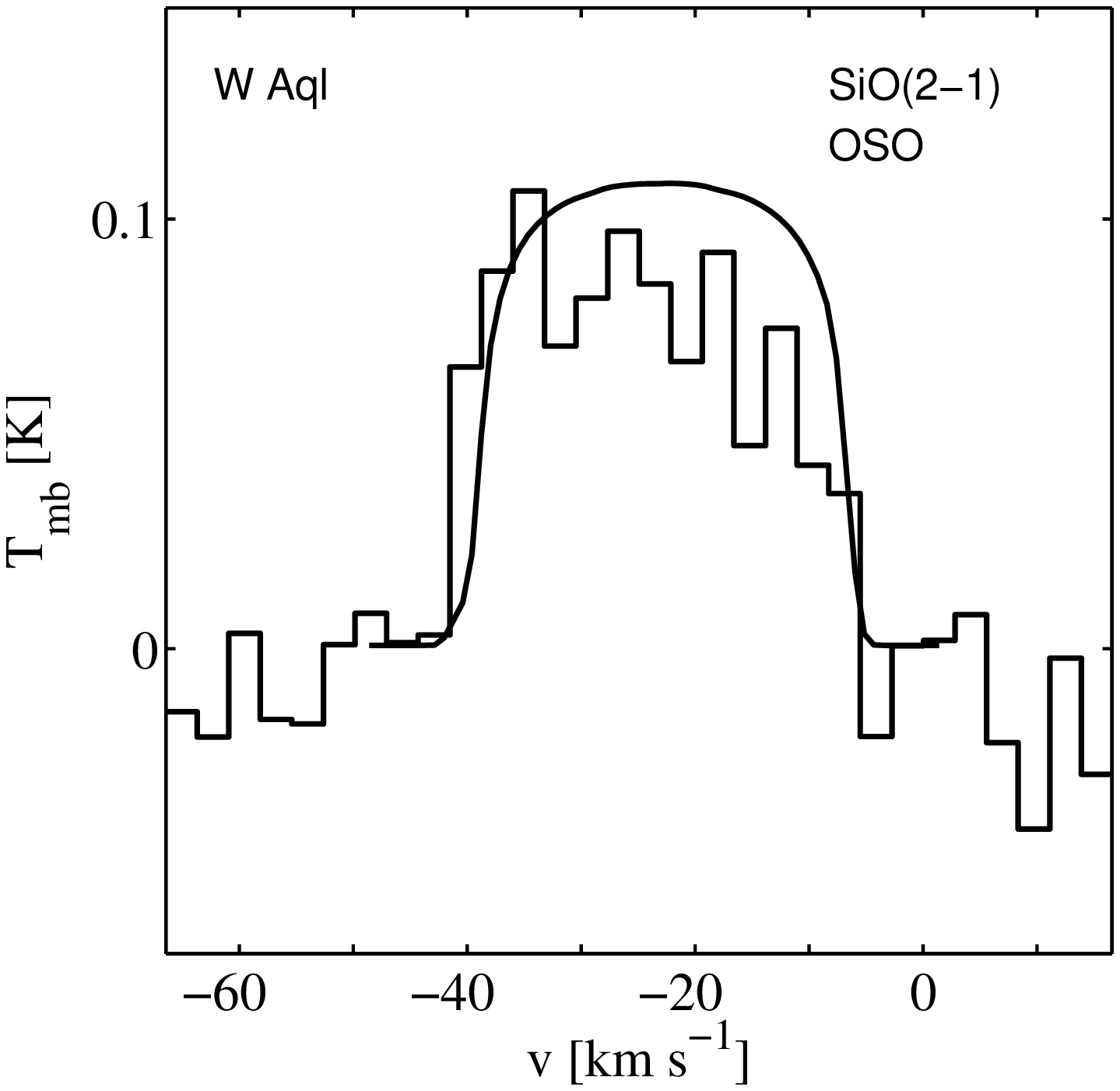}
\includegraphics[width=3.2cm]{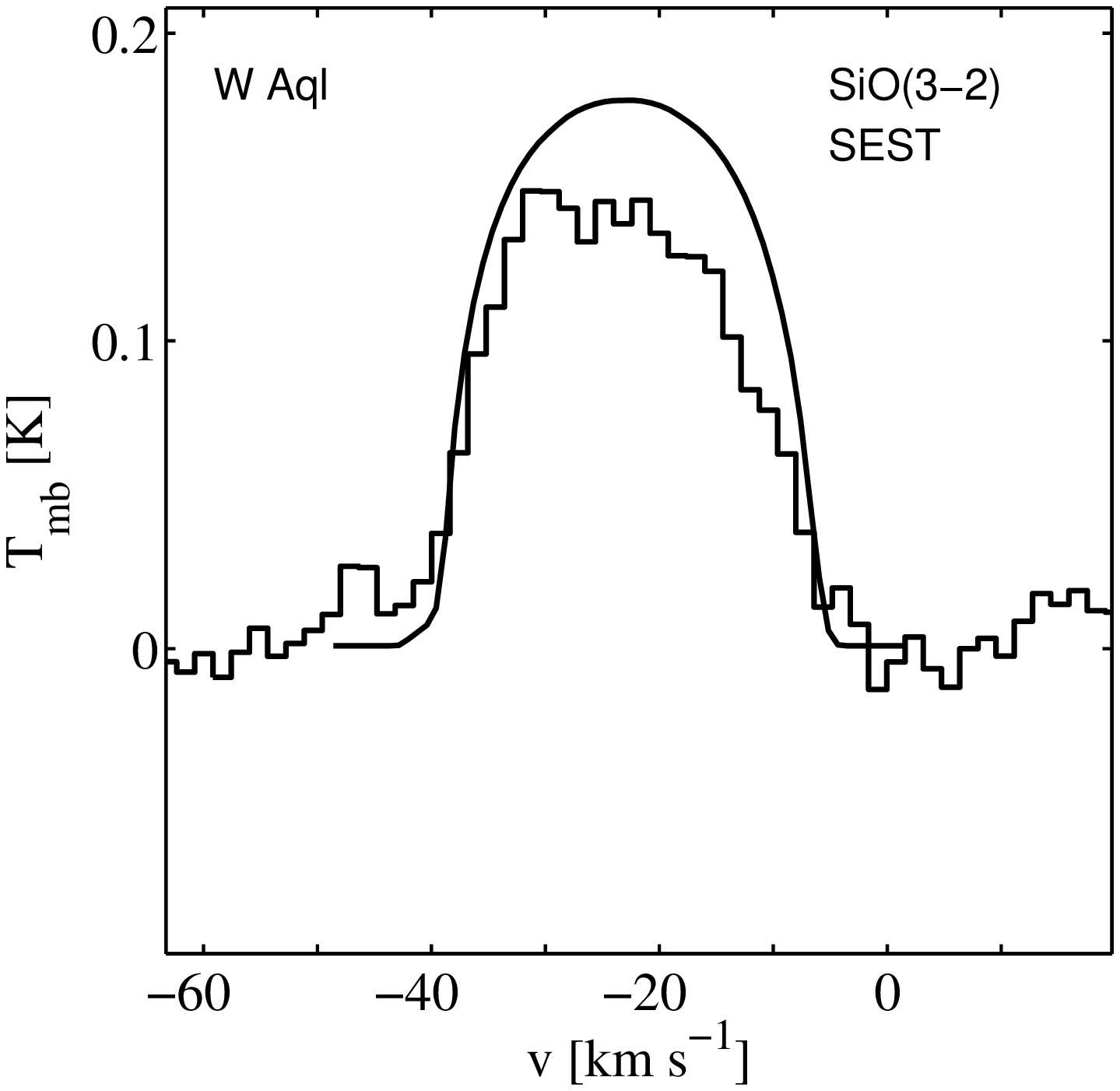}
\includegraphics[width=3.2cm]{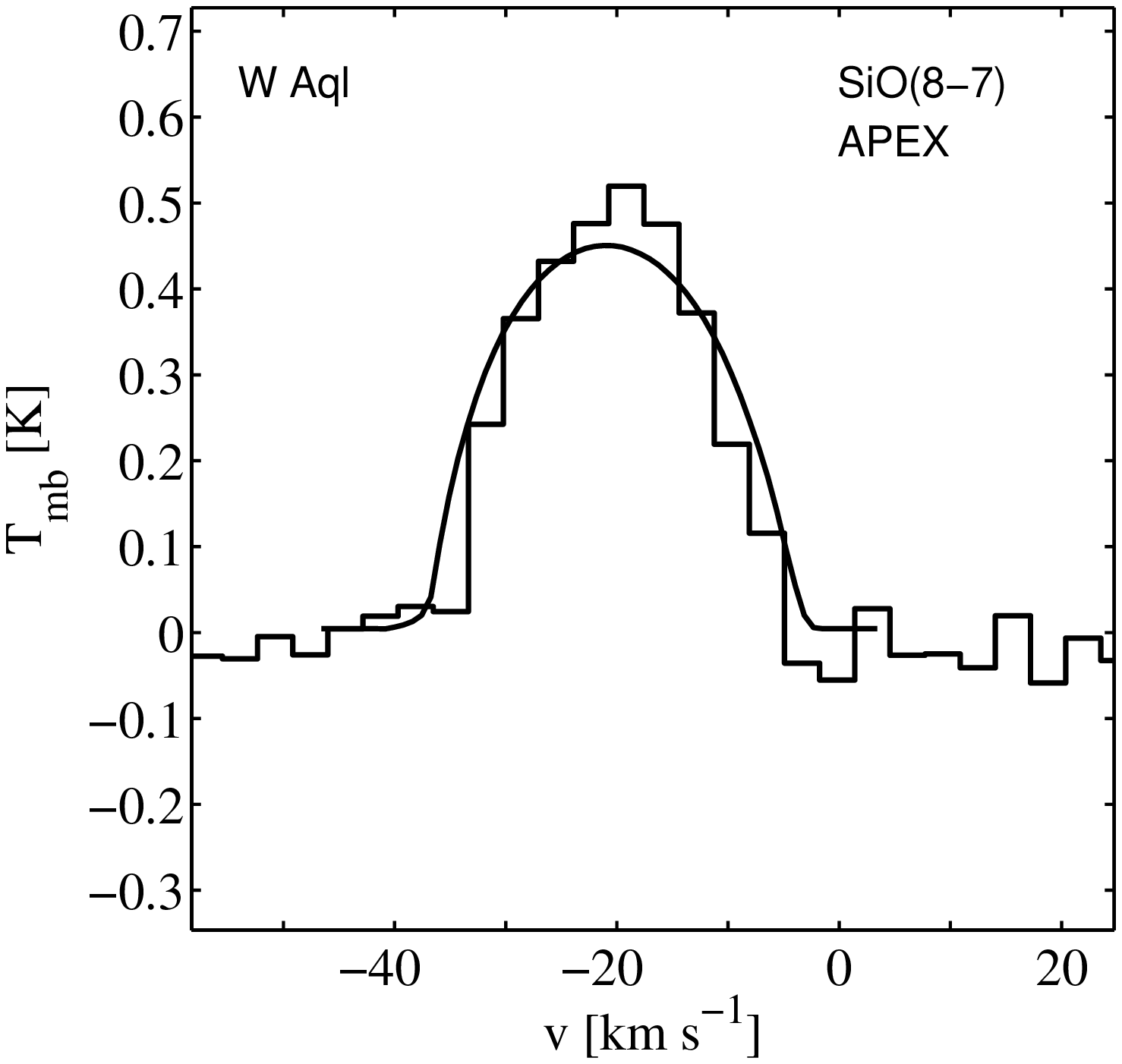}
\includegraphics[width=3.2cm]{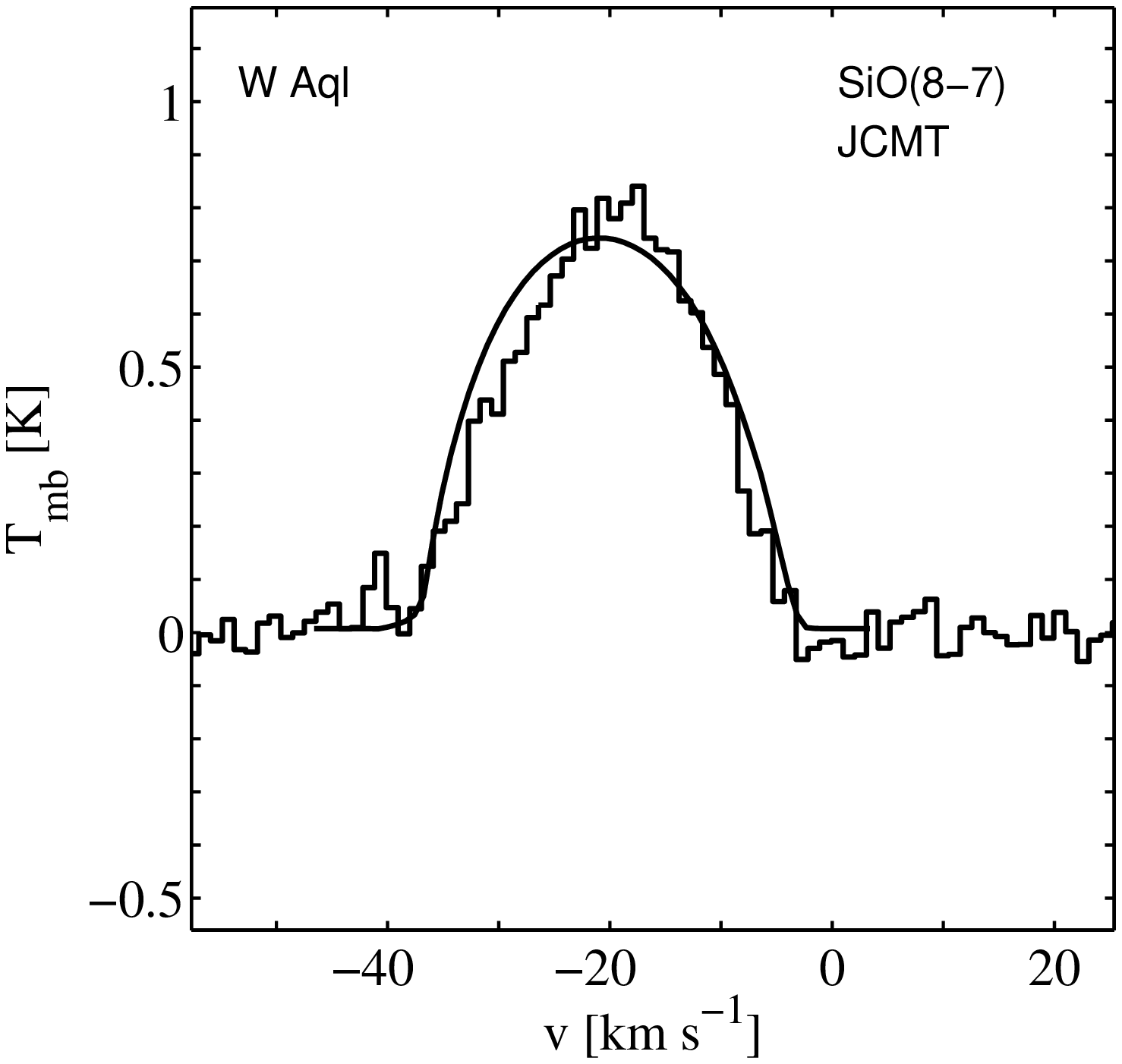}
\caption{Example of spectra (histogram) of several transitions of SiO from the S-type star W~Aql. The spectra are overlaid by the results from the best-fit model (solid line) for this source, assuming an initial fractional abundance of $f_{0}=3\times10^{-6}$ and a e-folding radius of the Gaussian SiO abundance distribution of $r_{\rm{e}}=6.5\times10^{15}$. }
\label{sio}
\end{flushleft}
\end{figure}

\section{Results and comparison with previous results for M-type and carbon stars}

\subsection{CO}
The median mass-loss rate found for the S-type stars is 2.7$\times$10$^{-7}$\,M$_{\odot}$\,yr$^{-1}$, and the distribution is spread over about two order of magnitude. The median gas expansion velocity is 8.0\,km\,s$^{-1}$, ranging from 3 to 21\,km\,s$^{-1}$. The mass-loss rate distribution looks very similar regardless of chemical type (Fig.~\ref{dist}). So does the distribution of expansion velocities, however, the carbon stars seem to have slightly higher expansion velocities. The correlation between $\dot{M}$ and pulsational period, and expansion velocity and pulsational period, also appears to be similar regardless of the chemical type. All these results point to that the mass-loss is driven by the same mechanism or mechanisms in all three chemical types. For further details on the CO results, see \citet{ramsetal06,ramsetal09}.

\subsection{SiO}
The SiO fractional abundances for the S-type stars range three orders of magnitude, and the median value, 6$\times$10$^{-6}$, is almost one order of magnitude larger than what would be expected in chemical equilibrium \citep{cher06}. From a non-equilibrium model, including the effects of shocks in the circumstellar gas, the SiO abundance would be expected to be a few times 10$^{-5}$. This is more in agreement with what we find in our models. A comparison of the results for all three chemical types shows that the abundance distributions are very similar (see Fig.~\ref{dist}) and can vary up to two order of magnitude for a specific density. We interpret the large spread in abundances as indicative of the SiO chemistry being a consequence of the effects of shocks in the stellar atmospheres, as a shock chemistry would be very sensitive to other specific parameters of the star, like shock velocity for instance. For the M-type and carbon stars, previous results show a clear decrease in the SiO abundance with the circumstellar wind density, and this has been interpreted as an effect of SiO adsorption onto dust grains in a high-density wind. There is some indication of the same effect in the S-type stars, however, very few high-density S-type stars have been observed. For further details on the SiO results, see \citet{ramsetal09}.

\begin{figure}[t]
\begin{flushleft}
\includegraphics[width=4cm]{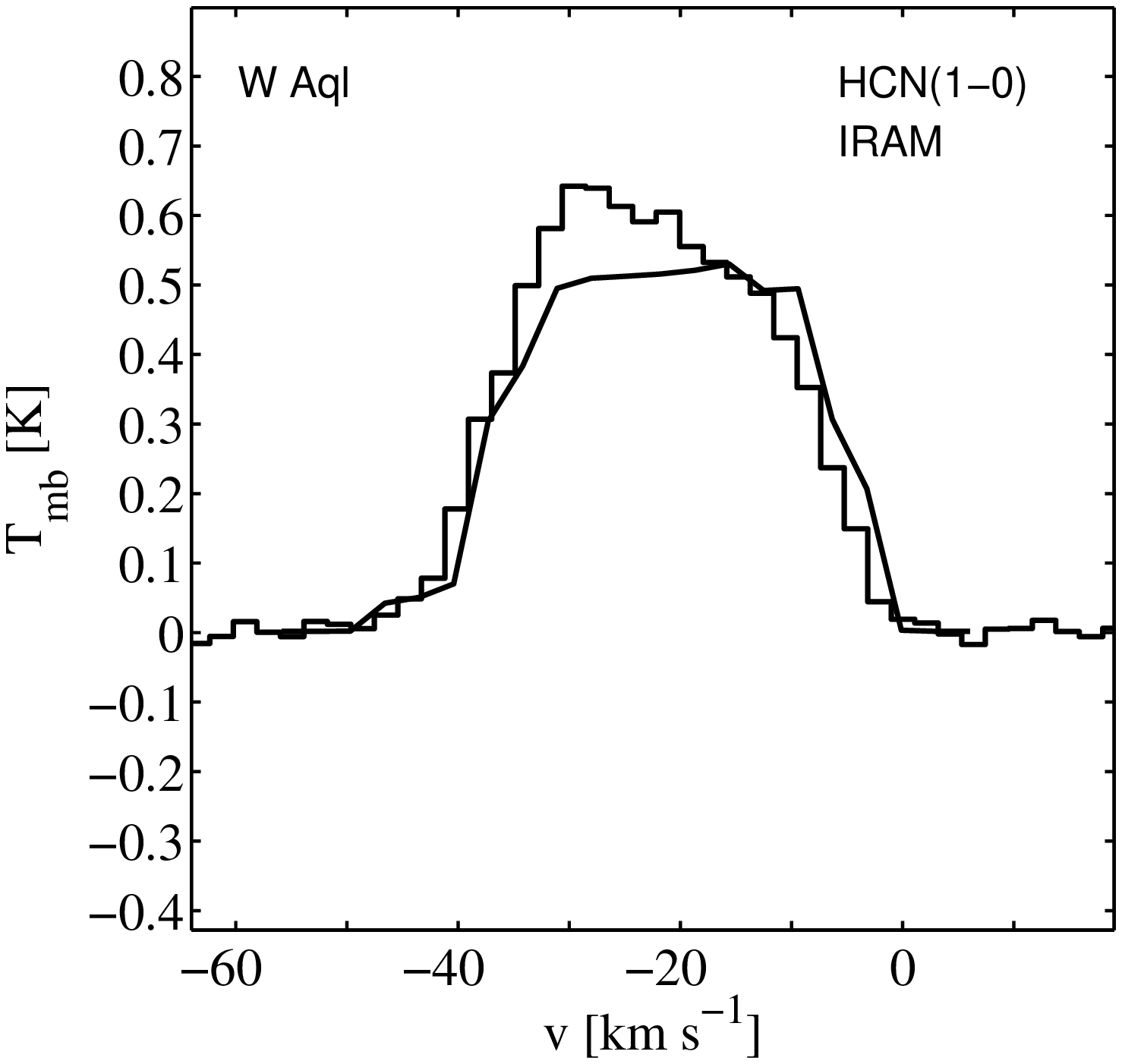}
\includegraphics[width=4cm]{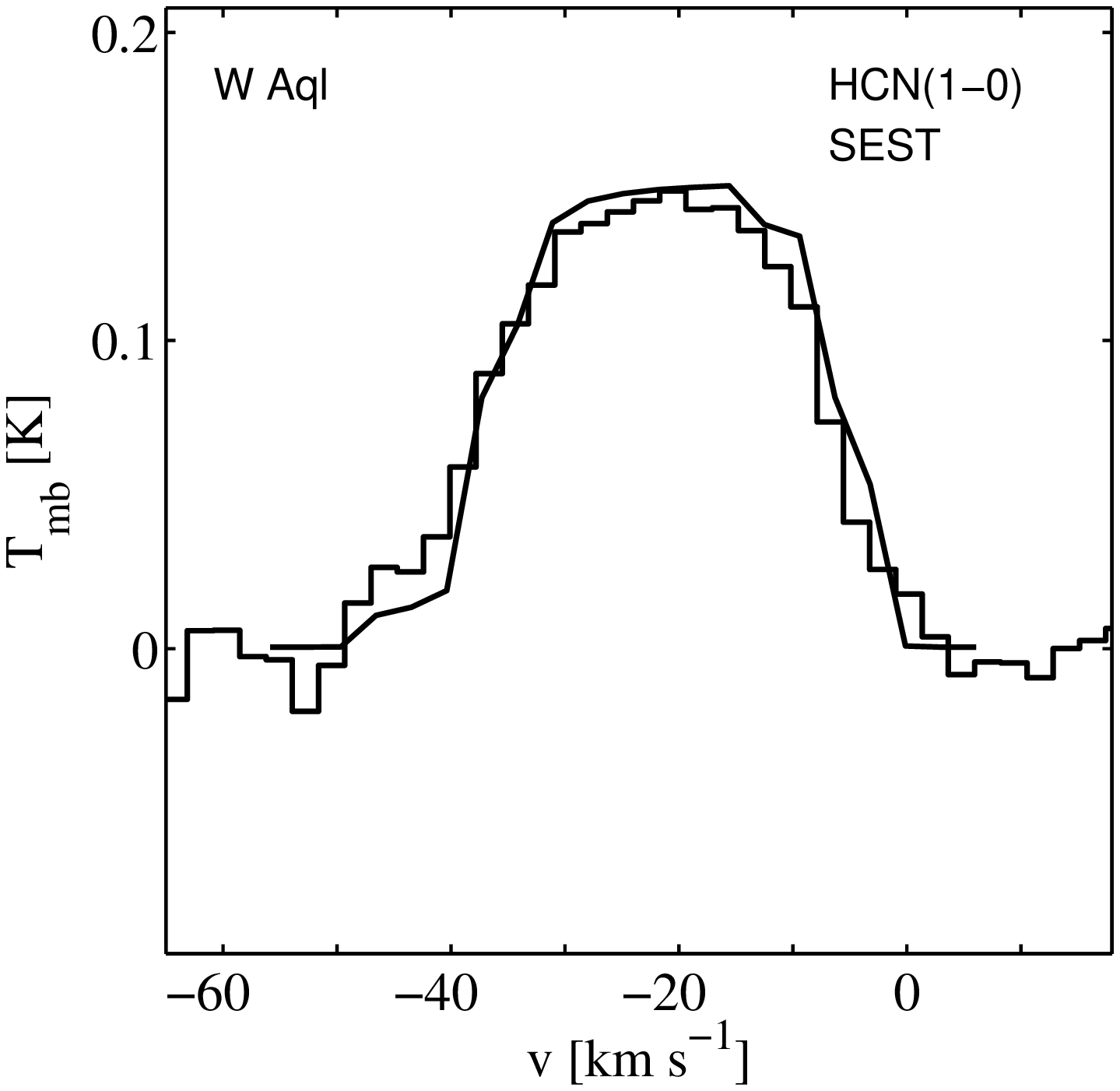}
\includegraphics[width=4cm]{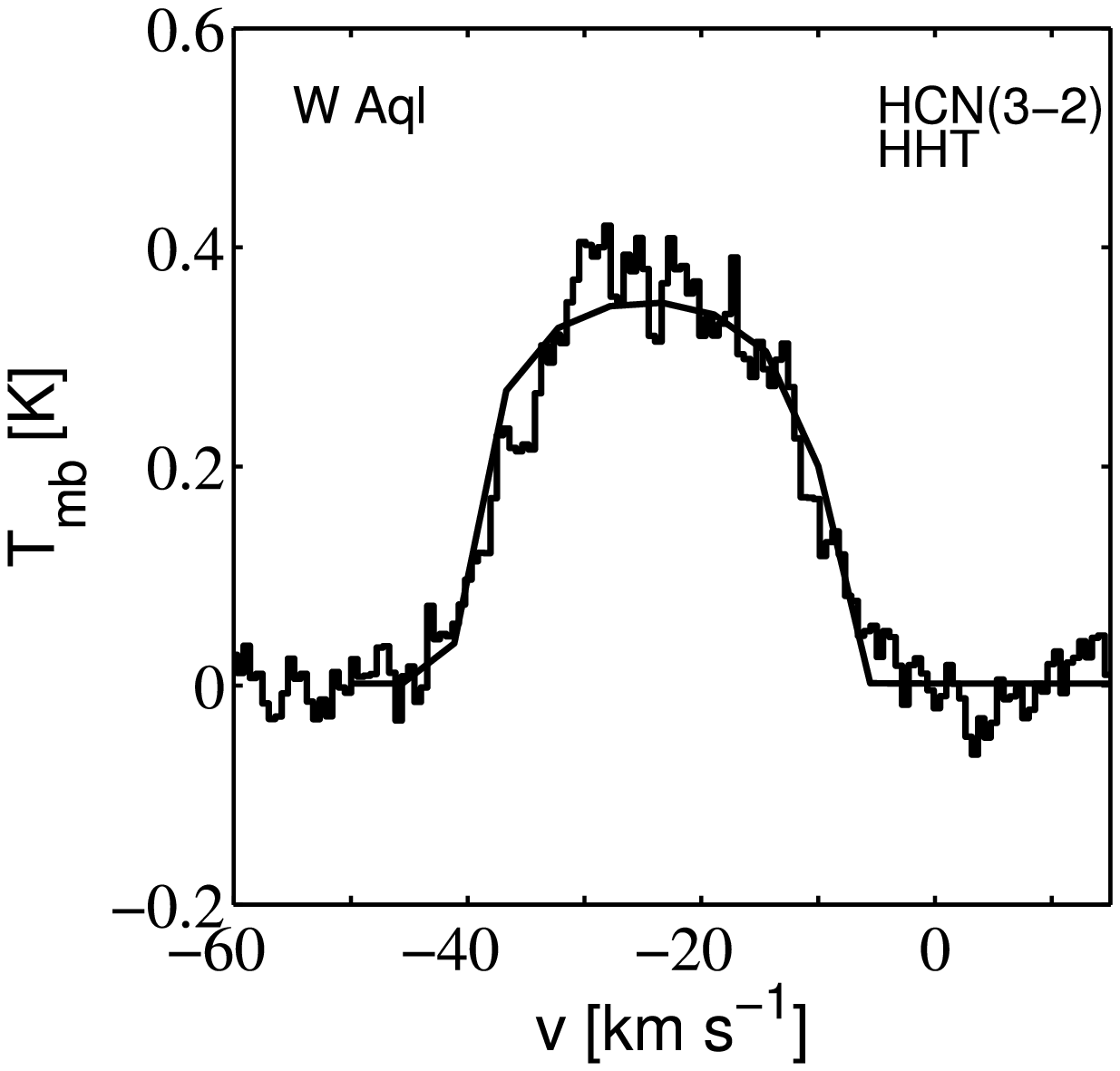}
\includegraphics[width=4cm]{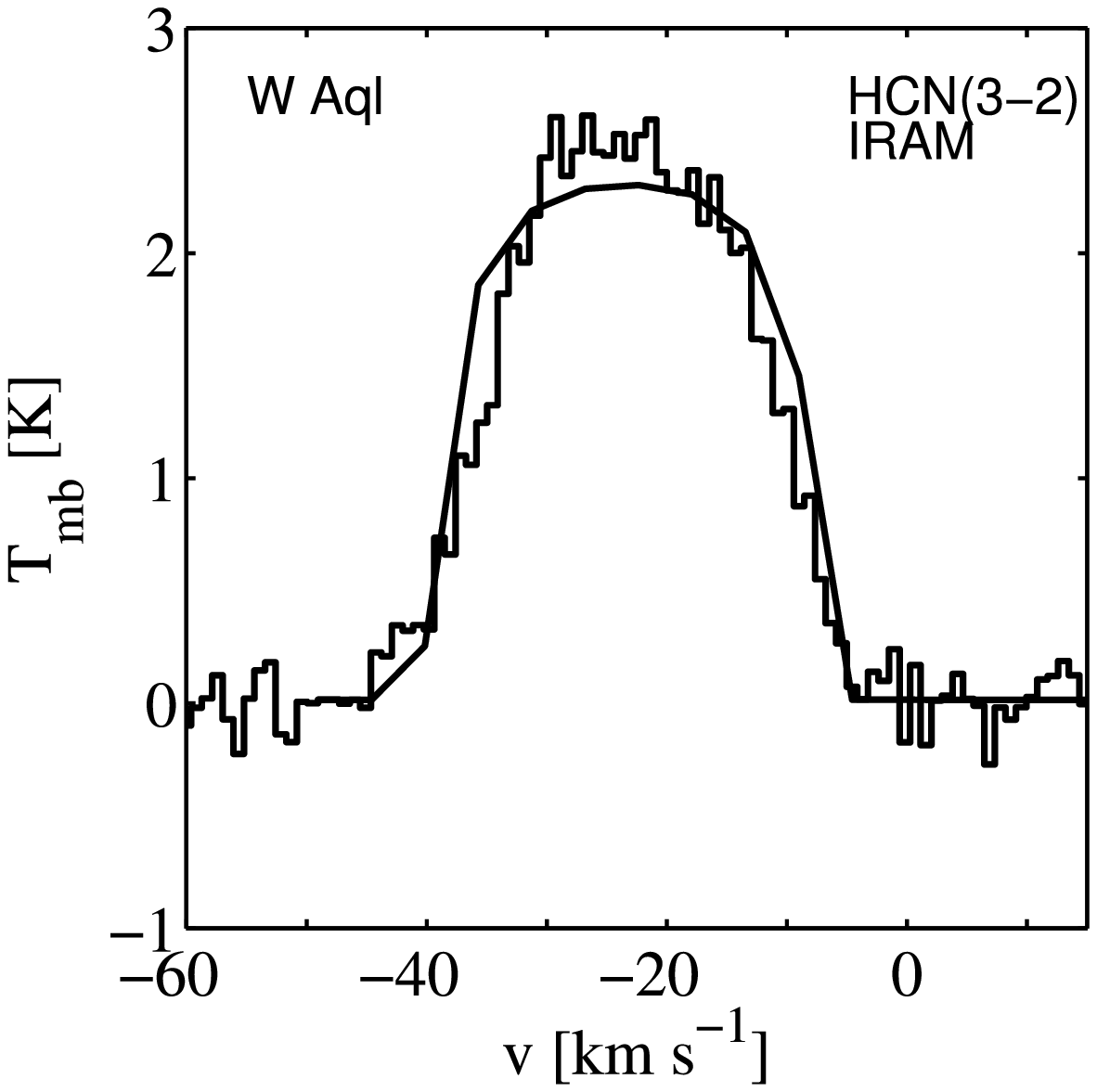}
\includegraphics[width=4.cm]{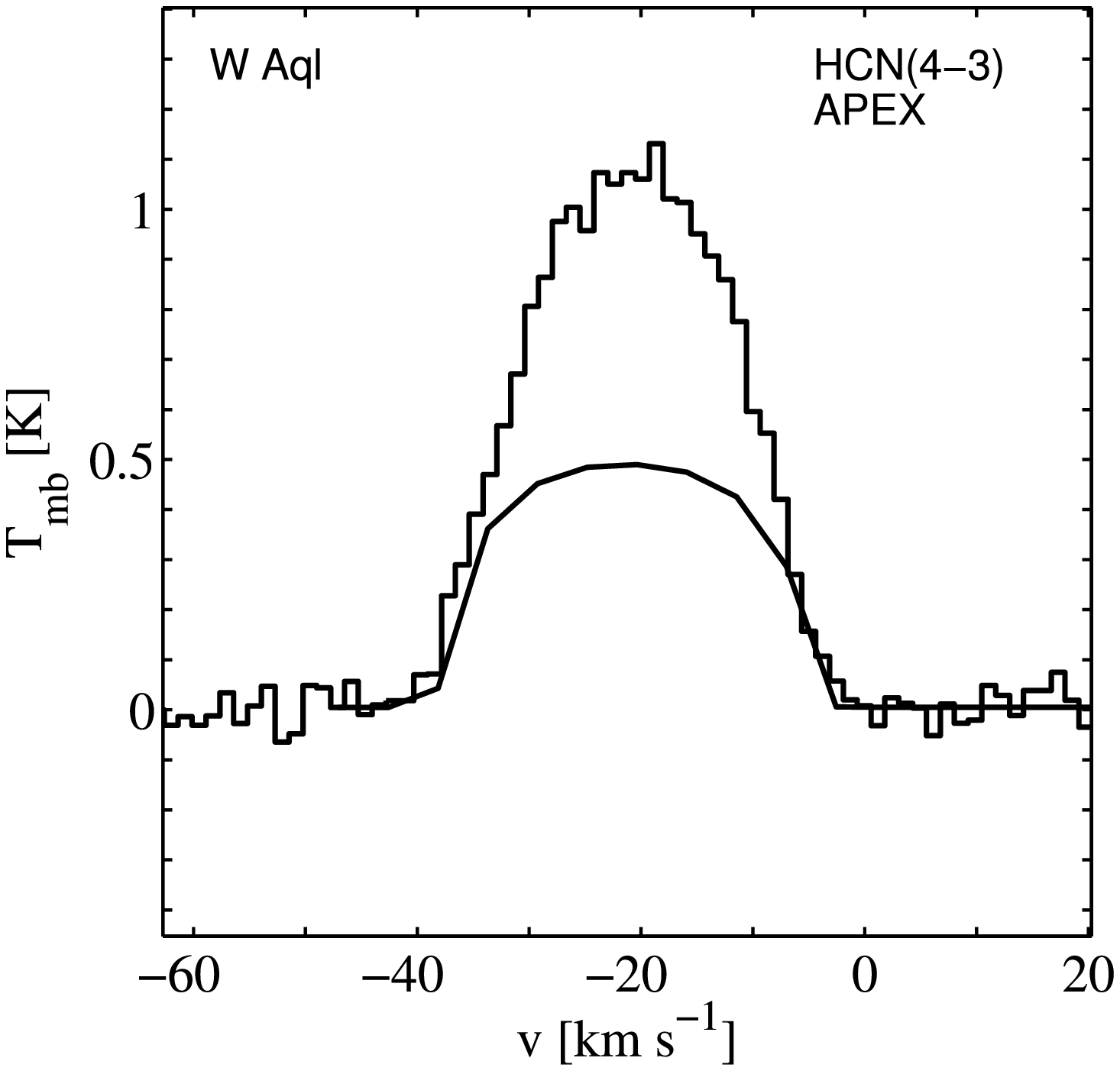}
\includegraphics[width=4cm]{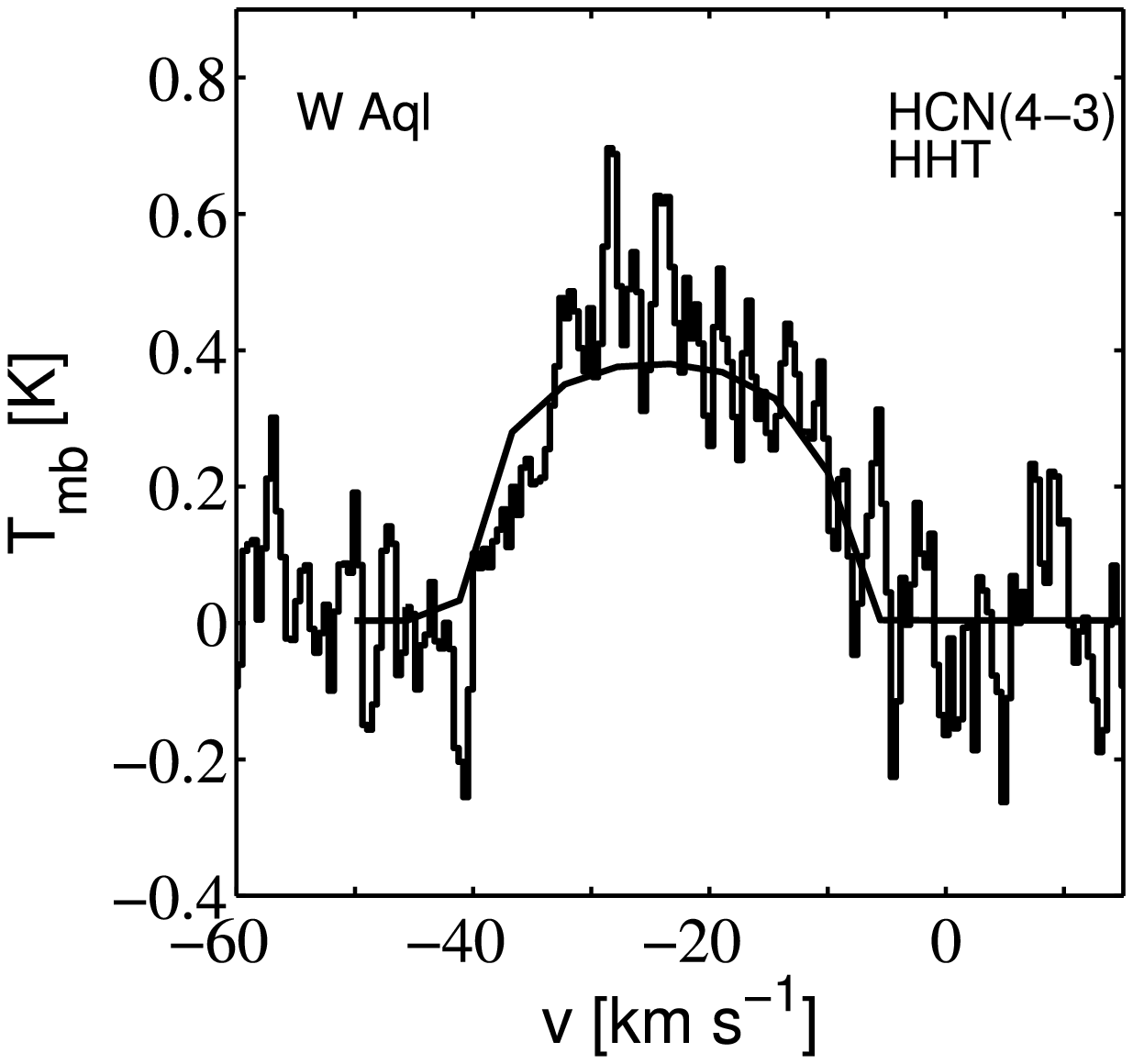}
\caption{Example of spectra (histogram) of several transitions of HCN from the S-type star W~Aql. The HHT spectra are from \citet{biegetal00}. The spectra are overlaid by the results from the best-fit model (solid line) for this source, assuming an initial fractional abundance of $f_{0}=5\times10^{-7}$ and a e-folding radius of the Gaussian SiO abundance distribution of $r_{\rm{e}}=6\times10^{16}$. }
\label{hcn}
\end{flushleft}
\end{figure}

\subsection{HCN}
\begin{figure}[ht]
\begin{center}
\includegraphics[width=4cm]{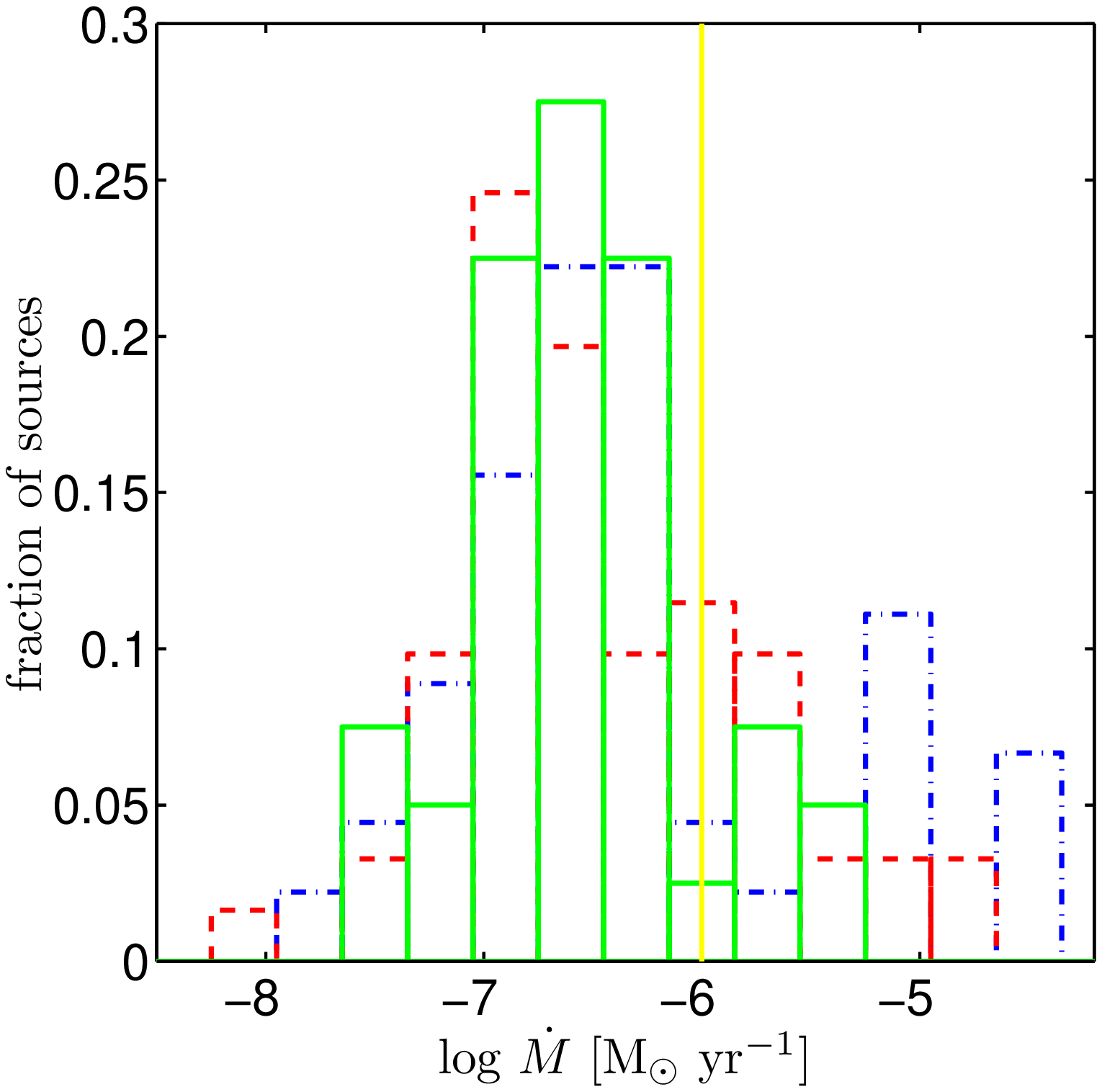}
\includegraphics[width=4cm]{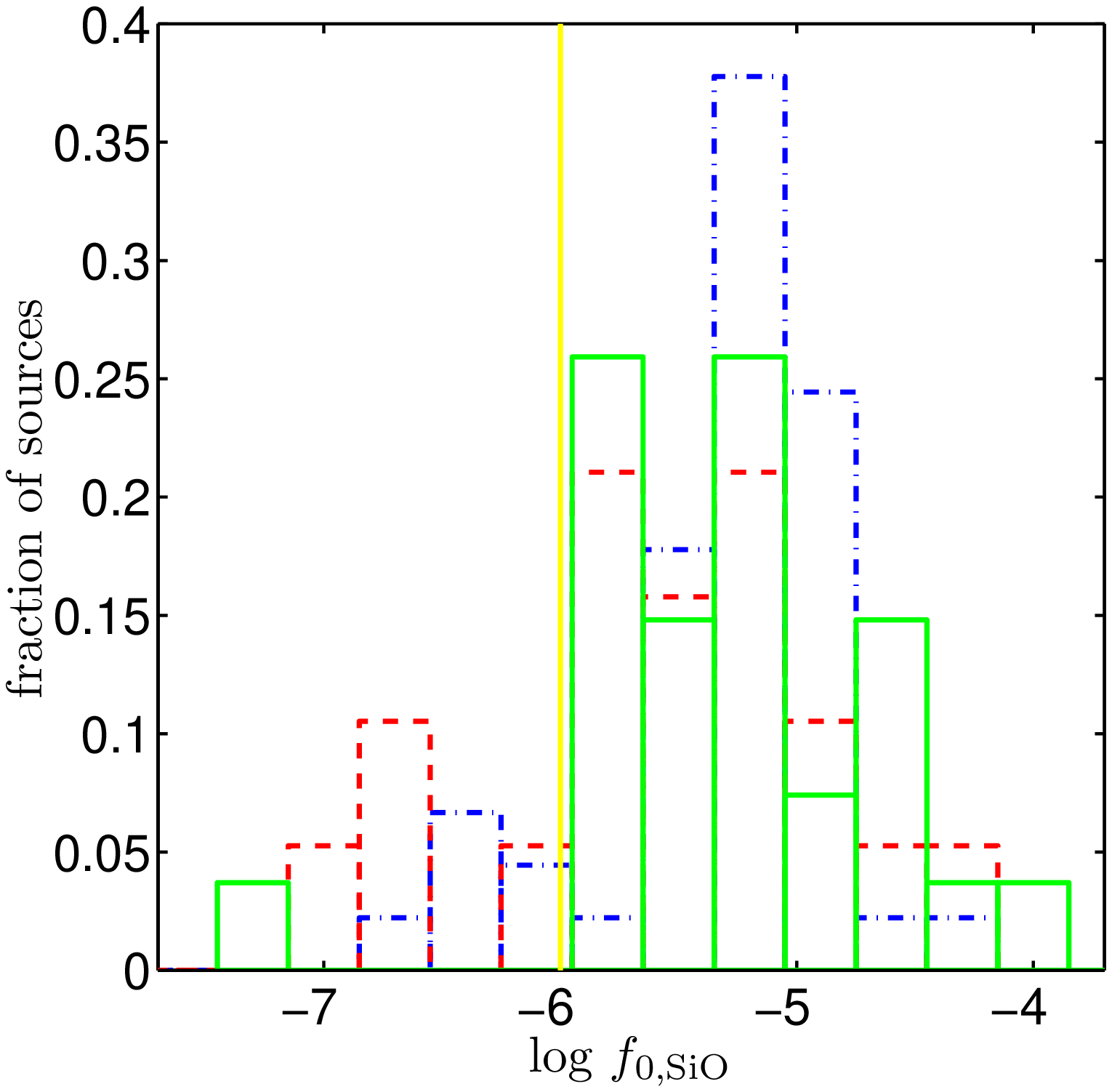}
\includegraphics[width=4cm]{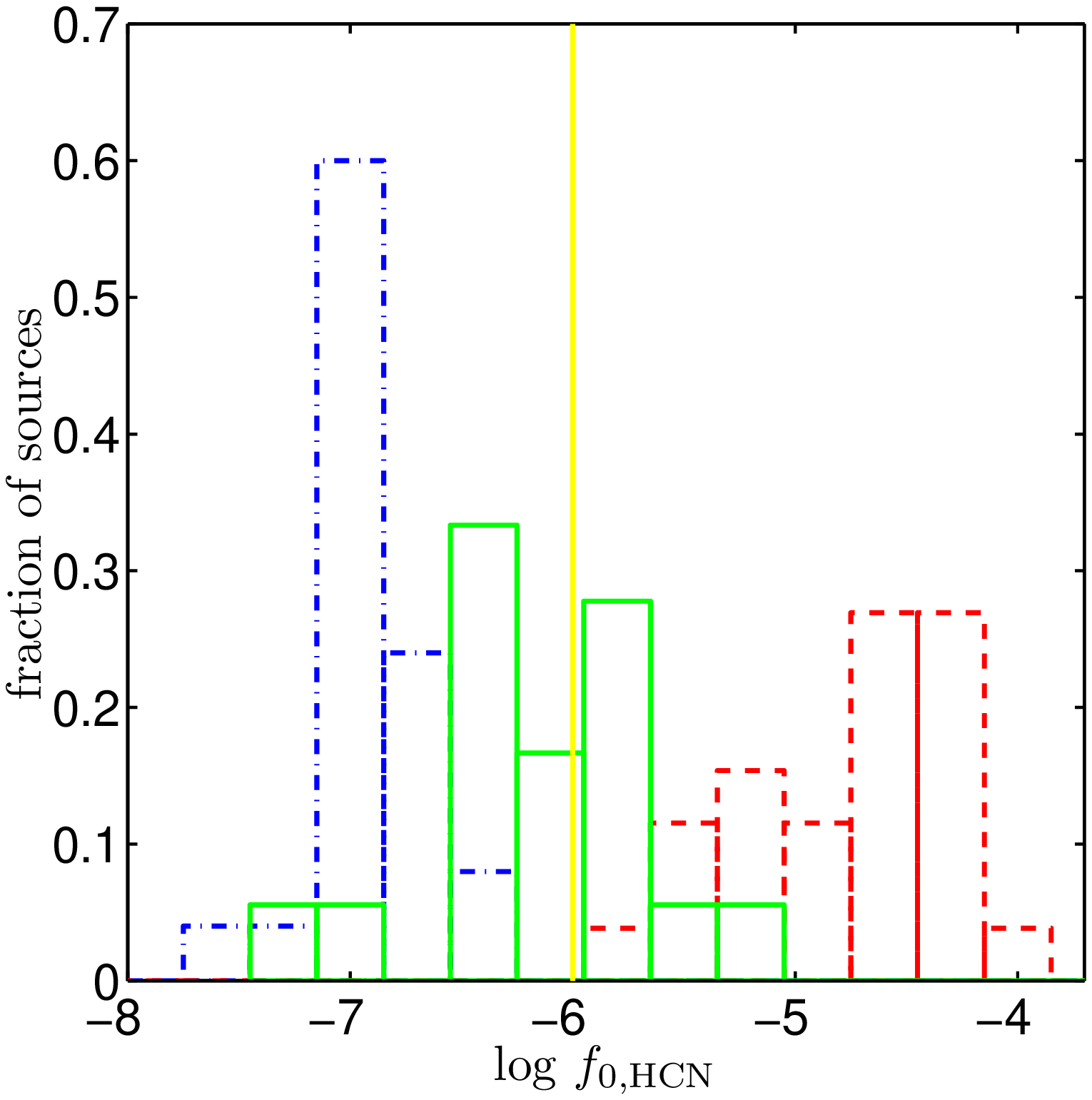}
\includegraphics[width=4cm]{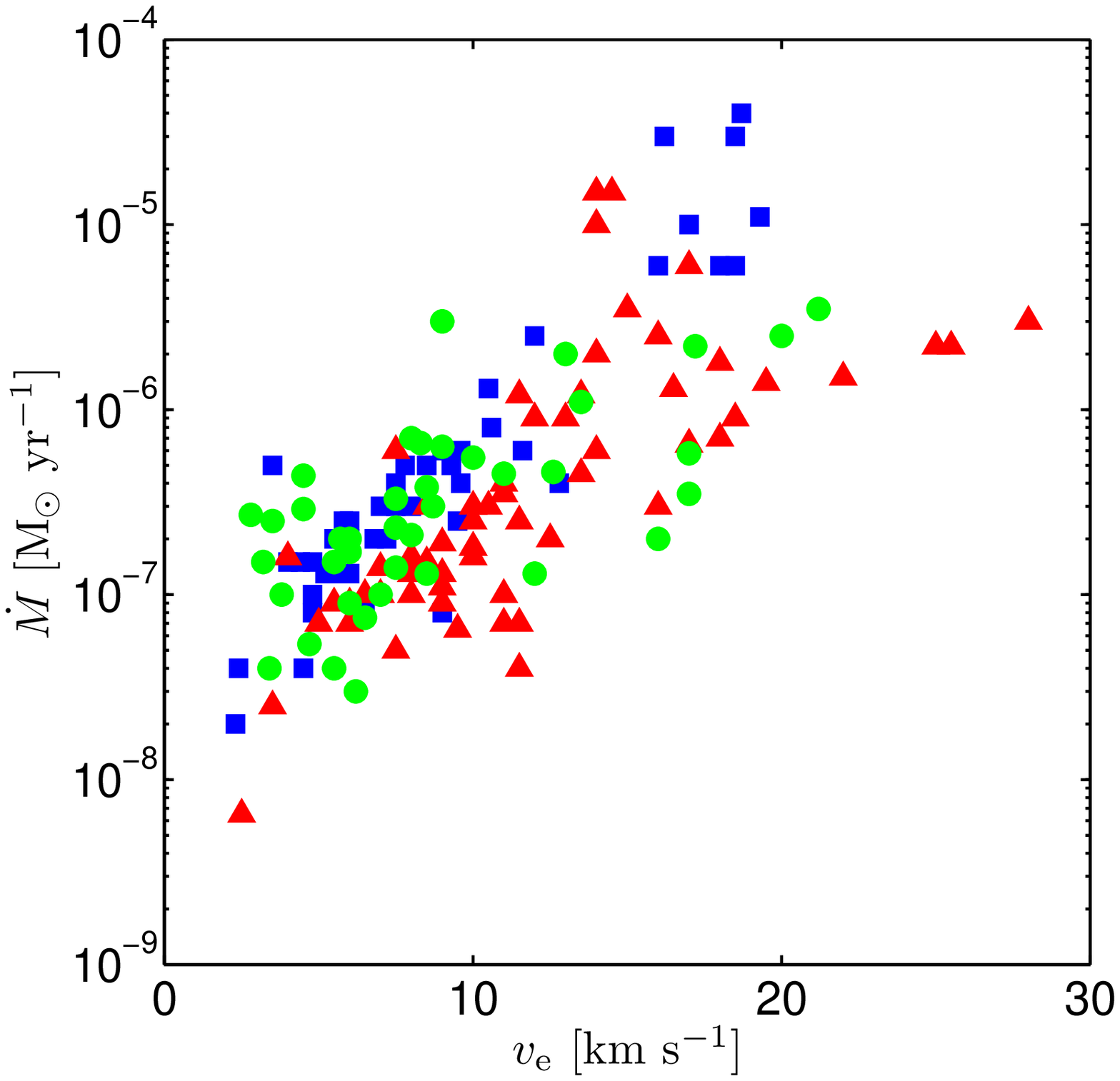}
\includegraphics[width=4cm]{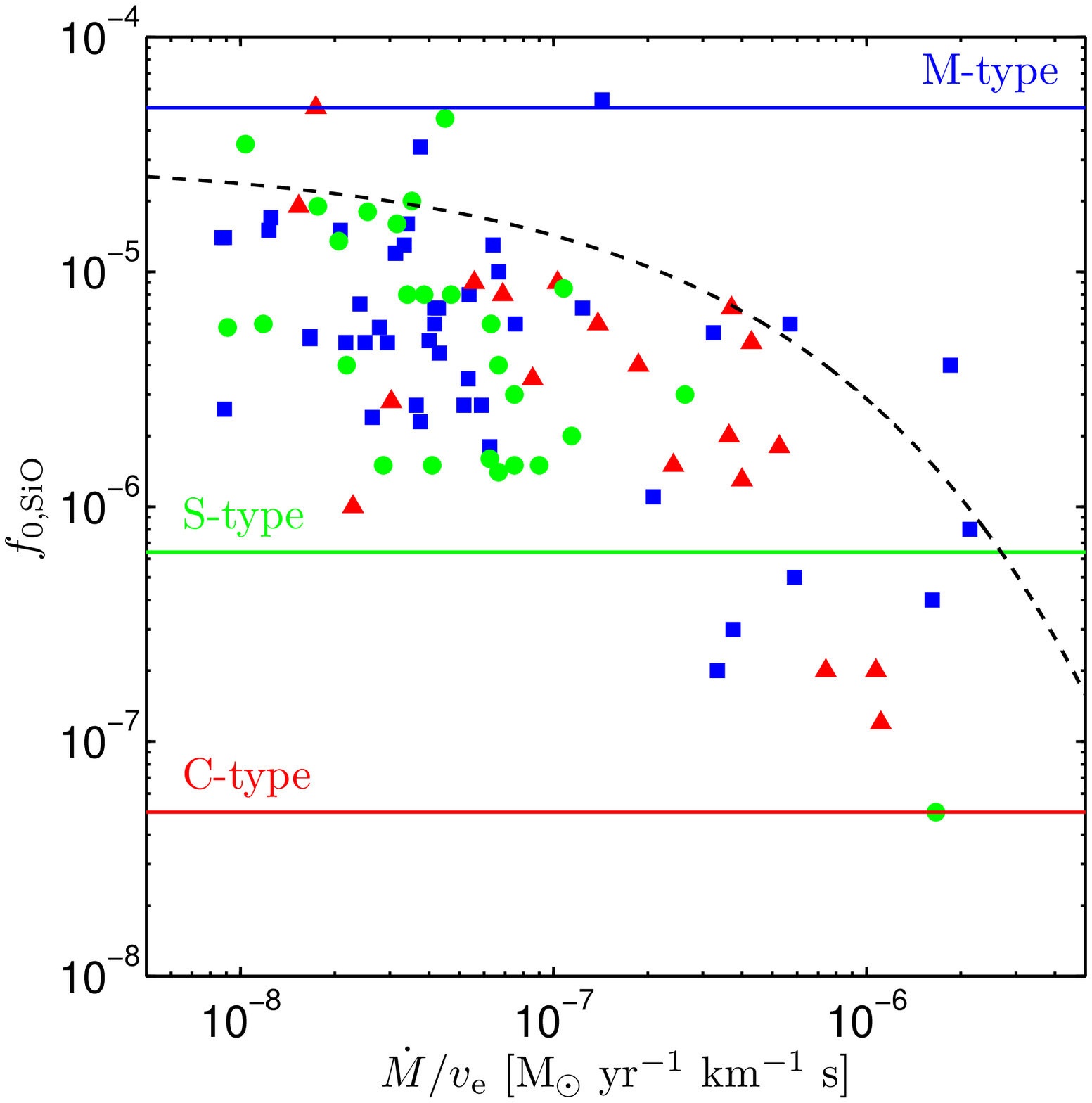}
\includegraphics[width=4cm]{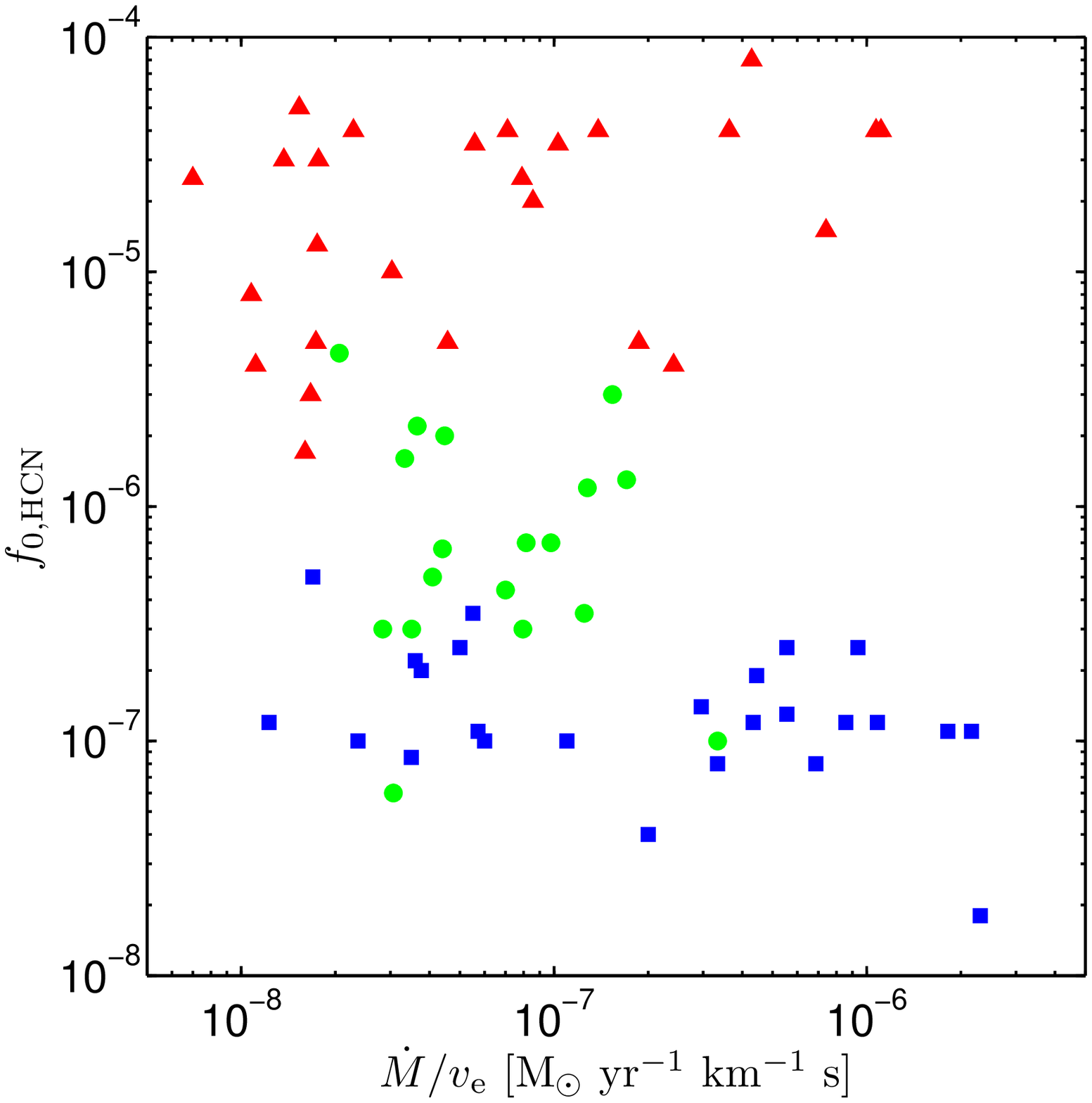}
\caption{The upper panel show histograms of the mass-loss rate distributions, the SiO abundance distributions, and the HCN abundance distributions of the three samples compared in our work. The solid line represents the S-type stars \citep{ramsetal09}, the dotted-dashed line represents the M-type stars \citep{olofetal02, delgetal03}, and the dashed line represents the carbon stars \citep{schoolof01}. The lower panel shows $\dot{M}$ versus $v_{\rm{e}}$ (left), the SiO abundance versus wind density (middle), and the HCN abundance versus wind density (right). S-type stars \citep{ramsetal09, schoetal10} are shown as dots, M-type stars \citep{delgetal03,schoetal10} as squares, and carbon stars \citep{schoetal06,schoetal10} as triangles. }
\label{dist}
\end{center}
\end{figure}

The HCN abundances show a different picture. Here the three types are clearly different. The carbon stars have a median abundance of 2.5$\times$10$^{-5}$, while the M-type stars have a median abundance of 1.2$\times$10$^{-7}$ and both distributions are quite narrow (see Fig.~\ref{dist}). The S-type stars, on the other hand, are spread out between the other two types, and has a median abundance of 7.0$\times$10$^{-7}$. This is more in line with what would be expected in equilibrium chemistry where the HCN abundance would be very dependent on the C/O-ratio. In this scenario, the S-type stars would either be more 'M-type-like', have a lower C/O-ratio and HCN abundance, or be more 'C-type-like', have a higher C/O-ratio and HCN abundance. The estimated HCN abundances do differ from what is found in equilibrium chemistry models, especially for the M-type sources where the equilibrium abundance would be expected to be $\sim$10$^{-11}$. This indicates that there might be some non-equilibrium processes influencing the chemistry, however not as much as in the models by \citet{cher06} where the HCN abundance is found to be independent of chemical type. For further details on the HCN results, see \citet{schoetal10}.

\section{Conclusions}
We have modeled circumstellar molecular line emission from CO, SiO, and HCN for 40 S-type AGB stars. We have compared the results to previous results for M-type and carbon stars, and arrive at the following conclusions:
\begin{itemize}
\item{We see no indications that the mass-loss process is different in the S-type stars compared to that of the M-type or carbon stars.}
\item{Circumstellar SiO abundances are similar in all three chemical types and the results are indicative of shock chemistry and grain adsorption.}
\item{Circumstellar HCN abundances are clearly sensitive to the spectral/chemical type of the star and more in line (than the estimated SiO abundances) with results from models assuming thermal equilibrium. Furthermore, our results for HCN clearly shows that the S-type stars in our sample are chemically different from the M-type stars.}
\end{itemize}
Despite the limitations due to the selection criteria of the compared samples, we believe that our conclusions apply to AGB stars in general.

\vspace{1cm}

\acknowledgements The authors acknowledge support from the Swedish Research Council. SR acknowledges support by the Deutsche Forschungsgemeinschaft (DFG) through the Emmy Noether Research grant VL 61/3-1. 

\bibliography{ramstedt}

\end{document}